# Testing And Hardening IoT Devices Against the Mirai Botnet

Christopher Kelly[1], Nikolaos Pitropakis[1], Sean McKeown[1], and Costas Lambrinoudakis[2]

[1]School of Computing, Edinburgh Napier University, Edinburgh, United Kingdom
[2]Department of Digital Systems, University of Piraeus, Greece
Email:*40204337@live.napier.ac.uk*, {*n.pitropakis,S.McKeown*}*@napier.ac.uk, clam@unipi.gr*

*Abstract*—A large majority of cheap Internet of Things (IoT) devices that arrive brand new, and are configured with out-of-the-box settings, are not being properly secured by the manufactures, and are vulnerable to existing malware lurking on the Internet. Among them is the Mirai botnet which has had its source code leaked to the world, allowing any malicious actor to configure and unleash it. A combination of software assets not being utilised safely and effectively are exposing consumers to a full compromise. We configured and attacked 4 different IoT devices using the Mirai libraries. Our experiments concluded that three out of the four devices were vulnerable to the Mirai malware and became infected when deployed using their default configuration. This demonstrates that the original security configurations are not sufficient to provide acceptable levels of protection for consumers, leaving their devices exposed and vulnerable. By analysing the Mirai libraries and its attack vectors, we were able to determine appropriate device configuration countermeasures to harden the devices against this botnet, which were successfully validated through experimentation.

## I. INTRODUCTION

The Internet of Things (IoT) has become more popular than ever with the heightened thirst for making our planet *smart*. The ability to have electronics take over the jobs that people have no desire to do has become more attractive than ever. This advancement, however, is far more sophisticated than we can imagine and with it came new waves of cyber-attacks. A report from F5 labs found a 280% growth in attacks on IoT devices *with a large chunk of this growth stemming from the Mirai malware* [1] while Symantec's report [2] for the year 2016–2017 showed a 600% growth of IOT cyber-attacks.

Cyber-attacks can originate from a vast number of different sources, using a variety of methods. Cyber-physical attacks are a result of a breach to the sensors of an IoT device which enables the adversary to access the device's data, resources or even create a backdoor to the system which can go unseen. Software attacks originate from malware being installed onto a device which can corrupt or steal data, which could potentially be very valuable. A slightly more dangerous attack occurs when the device's encryption keys are compromised which allows the attackers full control over the system. Malicious parties can, however, cause serious damage to such devices and networks without acquiring actual physical access to them. Network attacks are performed when adversaries can insert themselves between the device and network to perform actions, such as redirecting packets or igniting a significant disruption, such as a Denial of Service (DDoS) attack. These DDoS attacks can be performed by a large collection of compromised devices, controlled by one master device, commonly known as a controller, that flood the bandwidth of a chosen system with traffic. In October 2016 the source code for the Mirai malware was leaked and made public by an anonymous user. This was the most common malware to infect IoT devices and was specifically designed to target and exploit vulnerable devices.

The term *Internet of Things* can encompass any form of device that is connected to the Internet. It is, however, becoming more common for the definition to refer to sensors and other devices which autonomously communicate with each other to share information. IoT devices can benefit anybody, ranging from individual consumers to large industries. According to Statista [3], there were 23.14 billion connected IoT devices worldwide in 2018, which is projected to increase to 75.44 billion by 2025 [4]. With the number of connected devices expected to triple in the next six years, the IoT has the potential to have an enormous global impact. Additionally, Statista [3] estimate that by 2020 all forms of Business use for IoT devices will surpass 3.17 billion installed units, compared to just 1.5 billion in 2014.

The sudden rise of malware attacks on these devices is a serious concern for consumers and businesses alike. By studying how malware, in this case specifically the Mirai malware family, performs, behaves, and infects devices, it is possible to develop countermeasures and understanding in order to reduce the number of attacks which can be performed on companies and private networks. To build towards this, we tested the robustness of four different IoT devices against Mirai infection and investigated device configurations to increase the security and reduce their susceptibility to Mirai.

The contributions of our work are summarised as follows:
- The architecture of a test network, where a Coolead IP Camera, a Raspberry Pi, a Siricam and a Virtual Simulated IoT device are placed.
- The orchestration of attacks against the IoT devices using the Mirai botnet attack vectors.
- A robust configuration and evaluation of the defensive mechanisms implemented on chosen devices to suppress the threat of becoming infected by Mirai.

The rest of the paper is organised as follows: Section II provides technical background and related literature on the Internet of Things, while Section III briefly describes the methodology of the experiment and the respective results. Section IV describes the suggested defence mechanisms applied to our testbed to reduce the probability of being infected by Mirai. Section V draws the conclusions giving some pointers for future work.

## II. Background and Related Literature

### A. Internet of Things Devices

The different types of IoT devices available to purchase range from smart bike locks to smart home devices, such as fridges and plugs. IoT has become such a diverse market, with new mechanisms for engaging with our homes and devices becoming available regularly. The possibilities have become endless and there is almost a sense of dependency on these devices developing in recent years. IoT devices can be categorised into various archetypes, described below.

Wearable devices have become very popular and sought after, often offering both health and fitness tracking capabilities. The integration of heartbeat sensors and step counters have upgraded the fitness landscape and opened the doors for IoT technology to thrive on it. Devices such as smart rings have been designed which offer features such as contactless payments and Near Field Communication (NFC).

Smart home devices offer the ability to have almost all the devices in your life connected to the Internet. Everything ranging from clocks, door bells, cameras, heating systems, light bulbs and even window blinds. The most notable devices include the Nest smart thermostats [5] and the Amazon Echo [6]. Nest made it possible to control the heating of a household from anywhere in the world using a mobile phone, offering features such as timing control, temperature control and smart alerts. Amazon Echo is a smart speaker with seven microphones facilitating clear voice interaction, allowing users to ask questions or perform actions without having to touch a screen, keyboard, or mouse.

Tesla opened the world to the concept of the smart car, which offered voice recognition and took the word impossible out of the self-driving car theory, with tests happening all over the world to make this a reality. Recently the introduction of a wireless Internet connection in many cities made the concept of a smart city become a reality and now technology such as intelligent traffic light systems exist, all made possible through the Internet of Things.

In the cyber security community, there is a frequent joke which states the *S* in IoT stands for *Security*. It is far from obvious that there is no *S*. The dispersion and propagation of devices belonging to the Internet of Things has propelled the adoption of such wide ranges of Internet connected devices, facilitating far more automation and information exchange than was previously possible. However, as with any device accessible to the Internet, there is no such thing as a completely secure device or network. Malicious parties and cyber criminals lie in wait, with complete stealth, for a vulnerability or misconfiguration to present itself.

The rise of new malware families targeting IoT devices has only exacerbated existing problems with botnets, and has led to attackers exploiting such devices to the extent that they are referred to specifically as *Thingbots*. F5 Labs conducted a report [7], following the timeline of the Thingbots discovery. It traced the first instance back to March 2016, with the Remaiten malware, which attacked home routers over Telnet to launch DDoS attacks. With DDoS attacks still the most utilised attack, these Thingbots are being used to install Tor nodes and proxy servers, steal crypto currencies, hijack DNS servers and siphon credentials. The process usually begins with adversaries performing global scans searching for known open ports that allow an administrator shell, followed by infection.

IoT devices are not just targeted at random by all the malware in the wild, they are chosen. A litany of security flaws and vulnerabilities make devices in the Internet of Things prime targets for exploitation by malware. Services made accessible such as Telnet and secure shell (SSH) are manufactured with default passwords that is hardcoded in the firmware and the ability to change these passwords is either a difficult process for consumers or is entirely impossible for them. If the consumer did, however, manage to change the administrator password, other services offered by the device remain vulnerable to exploitation [8]. For example, in 2015 the FBI issued a warning, which is still relevant today, to disable Universal Plug n Play (UPnP) [9]. UPnP is a process where a nearby device connects itself automatically to a network with no username or password required, providing absolutely no authentication what so ever. Malware can use this protocol to gain access to more IoT devices on that network.

An IoT device consists of two major elements: hardware, and software. Hardware is less susceptible to becoming exploited as it is the physical components of the device and can have every aspect of its design tested to ensure it meets the expected requirements. Hardware also needs to be physically put into the hands of attackers to become exploited and therefore if, placed in a secure location, is far easier to protect. Software, however, is far more complex and responsible for a wide range of problems and dangers associated within IoT. A major flaw within software is the inability to update firmware versions on many devices leaving the firmware version outdated and far more vulnerable. Security and firmware updates not only provide up to date malware protection but can patch any form of weakness or security flaw within the firmware's code as they are discovered.

### B. Mirai Botnet

In August 2016, MalwareMustDie [10], a security research blog, discovered the malware that shook the world, the Mirai botnet. A report on the blog stated the botnet was created using ELF (Executable and Linkable Format) binaries which are very common on operating systems such as Linux. They discovered that Mirai targets SSH or Telnet protocols, exploiting the hardcoded and default credentials on many devices to access these services. Mirai also has the capability to launch brute force attacks using a targeted database of known credentials for specified brands and manufacturers of IoT devices. The ELF format used to create Mirai is implemented into the firmware of IoT devices, such as IP cameras, routers and smart devices, which is what makes them so vulnerable to attack [11]. Mirai has very complex and intelligent code which allows it to defend itself once it has compromised a device, preventing other malware from infecting the device its embedded in. Thus, Mirai can remain in a device without the possibility of another malware interfering with it [12].

The effect Mirai has on devices it has infected has been studied in order to determine its effects on device usability, power consumption, and network bandwidth consumption [13]. On several smart devices the power consumed by Mirai was found to have a negligible effect on the device during normal operations, but when the device participates in an attack the network bandwidth used increases by a considerable amount. As for its impact on regular device operation, functions were delayed by milliseconds, but some attacks caused the device to crash and reboot there affecting and degrading user experience. A separate study [14] used HoneyPots to observe and monitor network behaviour on a network comprised of many IoT devices, analysing traffic from the likes of Bashlite and Mirai over a substantial timeframe. It received over 2,385,460 commands from IP addresses and its overall conclusion was that Mirai and Bashlite both evolved significantly over the experimental duration, which is a serious cause for concern. It was apparent that newer botnets were learning and becoming more effective at deploying their attacks and choosing what attacks were most destructive for different software and hardware configurations [14].

*C. Related Literature*

Kolias et al. [15] demonstrate that if confidential credentials are built-in to the devices firmware, but it lacks a channel to perform software updates, and the credentials leak, the information remains exposed for a substantial amount of time, thus making the devices vulnerable to multiple threats.

Pa et al. [16], analysing the rise of IoT compromises, found that a range of 43 distinct malware samples were capable of being executed on 11 different CPU architectures. Of these CPU architectures, all the CPU names used in Mirai's infection process were found linking the different types of malware to exploit the CPU chips used in IoT devices. The same work also investigates Telnet-based attacks, finding that 91% of the 29,844 scanned hosts were running Linux. A large percentage of the devices were found to be DVRs and IP cameras. The summary of the paper concluded that of the 39 days their IoTPOT was running, a set of sandbox environments running Linux OS for embedded devices with different CPU architectures, 70,230 unique hosts visited the device. Among them, 49,141 logged in and 16,934 attempted to download external malware binary files. This means over 20% of the Linux embedded devices were vulnerable to malware, which simulated a very similar firmware to those embedded in consumer IoT devices.

Angrishi et al. [17] claim that the Internet of Things is becoming the *Internet of Vulnerabilities*. Their work takes a closer look at the evolution of malware in IoT and how it has progressed up to the most predominant malware, Mirai. Findings show that a large percentage of malware infections recruit the IoT device into botnets, most commonly to perform DDoS attacks. Another common similarity is that they often brute forcing open terminal ports, such as Telnet and SSH, with default credentials. The authors also discuss the root causes of the vulnerabilities present in IoT devices and links it to the rush of implementing new products and services through 3rd parties.

Edwards et al. [18] whose work focused on the Hijime malware, a decentralised Internet worm for IoT devices, identified that the infection process was very similar to that of Mirai, using username/password combinations through Telnet or SSH ports to establish a connection. The worm also worked effectively on BusyBox embedded Linux devices. The Hijime was found to be using BitTorrent's DHT protocol for peer discovery and uTorrent's transport protocol for data exchange. This was a unique set of methods used compared to that of other malware including Mirai. The paper concluded that a difference of the Hijime worm compared to Mirai is that it offered no malicious payloads and was perhaps in a phase where the author wanted to expand its scale and accumulate a larger botnet before deploying attacking payloads.

Understanding the ports and protocols of interest is a key element in understanding many security vulnerabilities in the IoT. Members of the Honeynet Project [19] conducted research analysing 24 Internet attacks. From their work, they found particular protocols to be of interest causing vulnerabilities. Some of these were protocols exploited by Mirai including HTTP, Telnet and SSH. They discovered HTTP to be one of the main vectors for exploitation, with 245 HTTP packets received to their testing environment in the 24–hour timeframe. For the Telnet protocol, of 1,075 connections, 835 were able to login successfully with a username and password. 750 of these were generated by bots which were more than likely to be compromised IoT devices. The project continues to monitor its honeypots to study the attack vectors being deployed, however, a conclusion the authors came to was that these botnets rely on weak and non-existent security measures on their targets which IoT devices are providing.

To deal with the magnitude of devices being compromised there simply is not enough being done. Researchers at the institute for critical infrastructure technology [20] investigated the behavior of malwares such as Bashlite and Mirai in action as *Practise Runs* which implies they believe there is a much more severe and destructive wave of attacks coming at a level unseen. They believe Mirai demonstrates how negligently and rapidly developed IoT software has become and this will only fuel attackers to act upon it to leverage and devour it for malicious purposes. Their research has led the belief that the 3–minute delay between bots becoming re-infected after a reboot is not due to the malware taking time to rediscover its vulnerability, but because there are so many armies of botnets fighting to reclaim the vulnerable host first. A proposed approach to help wage war against attackers is the development of penetration testing tools for IoT software and hardware. This not only opens the possibility for experts in their designated fields to help discover vulnerabilities, it reduces manufactures costs by allowing others to discover the flaws for them. They also offer the suggestion of improvement at organisational level for security controls. This can include the basic training of cyber-hygiene and fundamental security controls which can limit the susceptibility to botnets.

As IoT continues to expand at its current rate, and its dependency escalates as society moves towards automation, there is a thriving need to lay the framework and set the foundations to ensure the security and safety is given the same fuel as the thirst for its functionality. Ahmad et al. [21] focused on large scale IoT security testing, benchmarking and certification states. They came to a conclusion that for the future of IoT, features and mechanisms need to be established and put in place to ensure that in large scale conditions, everything is duly certified and tested accordingly. Antonakakis et al. [22] provide a comprehensive analysis of Mirai's emergence and evolution, presenting the range of devices it targeted and infected, along with the attacks it executed. Their findings show that Mirai's emergence was primarily based on the absence of security best practices in the IoT space and not on the unique security challenges introduced by the IoT devices. More recently, Pitropakis et al. [23] created a whole framework namely *NEON* that collects information from multiple probes to win the fight against Advanced Persistent Threats and malware families that constantly mutate.

Our work differentiates from all the previous approaches because our target is twofold. First of all we want to test whether Mirai can still infect modern everyday devices using attack vectors described by the related literature. Additionally, we want to explore efficient and simple countermeasures that can improve the security of IoT devices and their robustness against the Mirai attacks.

III. EXPERIMENTAL DESIGN

Before diving into our experimental setup it should be noted that one of the key challenges of our work was understanding the Mirai malware, as it is a very complex piece of malware with a large number of functions and actions it could perform. Building on top of this, being able to infect the chosen IoT device and gain administrative privileges was our goal as different brands of device have different firmware configurations. As our work used a live and active piece of malware, it had to be contained properly, as it would risk leaking the malware onto the rest of our network, thus converting our infrastructure to a zombie army for Mirai. It is therefore a secure form of practise to use Virtual machines to conduct the testing on. There were two virtual machines used, one to act as a controller for Mirai, capturing reports once exploitation has taken place and to deploy the malware's code. The second virtual machine was connected to the router attached to the devices. The virtual machine was run from VMware Workstation 14 Player version 14.1.3 which is a free software package specialising in cloud computing and virtualisation to offer the use of multiple virtual machines running their own operating system independent of the host. The operating systems used to run Mirai in this experiment was Linux Debian 9.8.0.

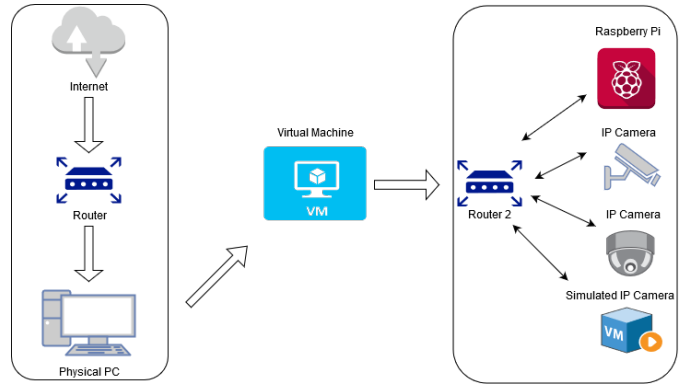

Fig. 1. Experimental Network Design

In recent cases of the Mirai malware being deployed onto devices, the source code has been known to delete the files and code used to infect the device after its execution, thus making it extremely difficult to find and examine the code [10]. The Mirai malware was designed to have at a minimum two servers to be able to effectively run the virus and manage to obtain a botnet to use. One server is needed to run the command and control aspect as well as a MySQL database to collect and store bot information. The second server is used to act as a scan receiver and a bot loader. The bot loader will use an advanced SYN scanner which is very resource efficient and will be used to execute the commands given to it from the command and control server once it has gained access to a bot.

The attack begins with scanning IP addresses to find potential victim devices to attempt to exploit. The source code of Mirai is interesting as it is hardcoded to avoid a specific set of IP addresses which can endanger the malicious parties, leading to their detection. The code is written to avoid scanning internal IPs so for this experiment the source code was edited so the code only scans IP addresses within our network. This is mainly to avoid infecting unaware people and preserve our resources. Another interesting aspect of the source code is Mirai's ability to assert its dominance by using techniques to remove other botnets and malware that might already be on the device, and then block any future malware from being able to cohabitate with it, effectively achieving a similar goal to the one in this paper.

A technique known as memory scraping can then take place, which removes all other botnet processes on the device's memory. This will actively look for QBOT [24], UPX [25] and Zollard [26] malwares which are popular viruses/worms on

IoT devices. QBOT is another botnet malware like Mirai, but is much less sophisticated. UPX is a software packer that can make reverse engineering of malware very difficult, requiring specialist knowledge. The Zollard worm is used for mining cryptocurrency on a compromised device without the user having any idea it is occurring. Another kill script was found in the code which eradicates a very similar botnet malware found in IoT devices named Anime/Kami [22]. The function is specifically designed for this malware and will instantly remove it from the device if discovered. Once IoT devices have been discovered through the IP address scan the next step Mirai takes is to attempt to remotely access it by logging into the device. Mirai performs a type of brute force dictionary attack by using a list of default credentials hardcoded into irresponsibly secured IoT devices.

Our experimentation was designed to be broken into four individual stages. For each stage to be a success it relies on the previous stage to have been accurately carried out with no problems and the objectives completed effectively.

- **Stage one** involves the setup of the network and devices. The network will be setup through the two virtual machines running and ready to serve commands. The chosen IP cameras will be setup according to the instructions provided by the manufacturer and no further changes will be made aside from the instructions given. These devices can then be setup to be discovered by the command and control server for Mirai on the virtual machine by setting up a chosen IP address block. An IP scan will then take place to ensure the devices are discovered on the network as intended and that they are functioning correctly.
- **Stage two** involves the setup of the Mirai malware. This will include any editing to the code required and a setup so the control and command server on the first virtual machine can begin the infection process. Next the bot scanner will be initialised to scan and gain access to the devices. In this stage the devices will become exposed to the malware. A time limit of ten minutes will be given which is more than enough time for Mirai to carry out its infection and will allow room for error should there be connectivity issues or any other networking issue. After exposure a full analysis of the results will take place.
- **Stage three** is the design and implementation of security measures based on the post exposure analysis. The analysis of the devices after stage two will expose the vulnerabilities in the devices and hopefully show what routes the malware took to gain access into the devices. This then allows rules and configurations to be designed and implemented onto the devices to help prevent them becoming compromised.
- **Stage four** will be a repetition of stage two, however this time the devices will have the newly designed security measures implemented on them. The aim of this stage is to discover how effective, if at all, the measures were and if they succeeded with their purpose to prevent exposure.

*A. Launching the attacks against our testbed*

Four target IoT devices were used for this experiment. Three of them are consumer products while the fourth was a virtualised simulation of an IoT device. We also developed a script to automate the installation of the Mirai malware on to the virtual machine. By running the script, it would automatically install all the necessary dependencies, cross compilers and functions to run the malware. However, it must be noted certain files still needed to be edited manually to include the IP address of the virtual machines as this was defined by the user and could not be predicted.

**Coolead IP Camera:** This device offers integrated Wi-Fi, local video storage and an FTP/mail alert system for its motion detection feature. To discover what ports where open on the IP camera an Nmap scan was done on its IP address which revealed it had the Telnet port 23 open, the HTTP port 80 and Asterix port 8600. To access the devices camera application, someone can visit its IP address in the browser on port 80 which prompted a username and password. To gain root access to the device and Telnet port a hydra attack was executed to attempt to crack the credentials. Based on the device not being properly branded, it was assumed default root credentials were hardcoded so a text file containing known root credentials could be used. After a successful login using the cracked credentials to Telnet it was discovered that BusyBox was installed and with this the version and commands allowing access to the file system, which could be cancerous if exploited by the likes of Mirai malware as it would have root privileges. The HTTP port was also accessed using the cracked credentials which gave full access to the cameras video feed and the option to fully customise all the settings including network configuration

**Raspberry Pi:** The operating system used was Raspbian Stretch which was connected to the network via ethernet. The raspberry Pi is becoming extremely popular around the world and allow developers and programmers to write their own software to use on this portable and simple device. The tool Nmap was once again used to discover what open ports the system came with by default. To SSH and Telnet into the device and to gain root access the default credentials were used which could be easily cracked if required. Access to these privileges would allow any form of tampering to the system and allow a malware like Mirai to hijack and turn this device into a bot without complications. To login and access the camera the HTTP browser was used and accessed by entering the devices IP address. The application simply needed the default credentials entered to allow access to the stream.

**Sricam IP Camera:** This camera offers a wall mount and metal shell to allow outdoor use and weatherproofing, with a 1080p HD capability, motions detection and Wi-Fi. The device uses an android, iOS or windows application, which requires the user to sign up and then discover the device on their network and add via a user name and password. An Nmap scan of the device only revealed two open ports which were unexpected. The device did not have a web server or Telnet server running and used ports used to stream its footage through the app. Port 554 is a Microsoft real time streaming protocol (RTSP) and is used to accept incoming and outgoing packets to a client streaming using the RTSP protocol. Port 5000 is used by Universal Plug N' Play to accept incoming connections from other UPnP devices.

The discovery that the device used RTSP lead to further research of the protocol to discover how it functioned and what role it played in allowing the cameras footage to be streamed to devices. As it was a real time streaming protocol run over port 554, there was a question of whether or not the stream could be intercepted and viewed in a media application without having to provide any form of authentication. It was discovered that the RTSP protocol was susceptible to a URL brute force attack by attempting to enumerate media URLs by testing for common paths. iSpyConnect [27] provided the endpoint of the URL for the Sricam model being used for the experiment. The media application used to intercept the stream was VLC player which allowed the ability to connect to a network stream using a URL. The application *Angry IP Scanner* was used to discover the IP cameras IP address. Typing the suggested address displayed the camera's live footage without having to provide any credentials or form of authentication.

**Simulated IoT Device (Ubuntu with Busybox):** The simulated device was setup using a VMware virtual machine on an isolated private network. The virtual machine was setup to run Ubuntu 18.04.1 LTS with BusyBox. This represented a vulnerable IoT device and was configured with default credentials, as well as running BusyBox's Telnetd server on port 23. The web server on port 80 was also configured to make it as similar to a physical IoT device as possible.

The first stage of our methodology was to setup the network and virtual machines. Two virtual machines running Debian 9.6.0 were installed to act as a command and control server and a bot scanner. The machines were setup using a bridged connection so that they acted almost entirely independent from the host machine. The IP addresses were allocated by DHCP in the local network by the router. Terminal was used to SSH into the virtual machines and login as a root user. This allowed the execution of commands, however, was not absolutely necessary as these could be executed in the virtual machine's terminal application. The application WinSCP was used as an FTP server to transfer the Mirai source code files onto the virtual machines and to edit any files that needed appending.

Soon afterwards, the Command and Control virtual machine was used to initiate the botnet. Once the botnet process was started, the CNC CLI panel was logged into using the servers IP address and the specified port that is chosen in the configuration. Once logged in the number of bots connected to the botnet as well as the range of attack vectors can be seen. The next step was to perform a zmap scan on the network to search for active IPs with the port's Telnet and/or SSH open. Once these were discovered the attack could take place to attempt to login to the newly made list of IP addresses. Mirai uses its pre-defined list of default credentials to attempt to authenticate and login to the devices. As previously discussed some of our devices are using root logins, thus making them extremely likely to be infected. Next the devices are brute forced in the attempt to login with root privileges.

If successful, the malware will use the tool *wget* to install its payload and binaries onto the device using commands that are pre-configured in the source code. If the device does not have wget installed, the malware has an alternative and attempts to use TFTP instead to infect the device. If the compromise and brute force attack is a success the loader will show the successful login in the text and on the botnets CLI panel the bot count will increase by one. Next, one by one, the devices were connected to the local network and exposed to the botnet process in an attempt to infect the device and in turn, convert to a bot. Confirmation of the malware successfully compromising and infecting the device is shown in the command and control panel with an increase in the bot count from *0* to *1*.

### B. Outcomes of the Attacks

**Coolead IP Camera:** As this device was found to be using Telnet running BusyBox and HTTP it was assumed Mirai would be successful in its attempt to compromise the device. This was further backed up by the cracking of its credentials which were found to be simple defaults that were included in Mirai's password list. The device was compromised in 9 seconds.

**Raspberry Pi:** The Pi was running both Telnet and SSH as well as HTTP. Documentation for the installation states that it comes pre-configured with its default credentials defined, and it is assumed a large majority of Raspberry Pi users, of which a large percentage are amateurs to the computing industry, would leave these credentials unchanged. The device was compromised in 11 seconds.

**Sricam IP Camera:** The Sricam IP camera proved itself to be a well-protected device in terms of the malware exploitation. Although it was found to leak the video footage without any credentials needed via a ONVIF network stream using VLC, its lack of open ports that allow a shell to perform terminal commands proved secure. The malware needs to be able to execute commands directly onto the device and the Sricams function to only use the essential ports to run its app was a fundamental reason this camera came out uncompromised.

**IoT Simulated Device (Ubuntu with Busybox):** As this device was simulated, it contained a very similar setup to the Coolead IP camera with its running Telnet server and BusyBox install. This, like the Coolead, suggested it was a prime and vulnerable target for the Mirai and would be compromised significantly easily. As it was designed to have default credentials, it was presumed confidently it would be easily cracked by the brute force and Mirai loader, thus being compromised in 9 seconds.

## IV. SUGGESTED DEFENCES

As the devices have been exposed and in certain cases, compromised, defensive measures are required in order to minimise the possibility of the Mirai malware being able to infect the devices again. An analysis of the modifications made by the malware on compromised devices was performed to see what changes had been made. This assisted us in the decisions of what measures need to be implemented. The proposed counter measures are summarised in Table I.

TABLE I
DEFENSIVE MEASURES PER DEVICE

| Defensive Action \\ Device | Coolead IP Camera | Raspberry Pi 3 | Sricam IP Camera | Simulated IoT Device |
|---|---|---|---|---|
| Change Telnet Credentials | Password changed | Telnet disabled. SSH made primary protocol | Telnet Not Enabled | Password changed |
| Change Telnet Port | Randomised port | Randomised port | SSH Not Enabled | Randomised port |
| Changed Web browser credentials | Password changed | Password changed | HTTP Not Enabled | Password changed |
| Disable SMTP | N/A | Protocol disabled without root privileges | Not Enabled | N/A |
| Replace HTTP with HTTPS | N/A | HTTPS with self signed SSL certificates | HTTP Not enabled | N/A |
| Busybox shell exclusive to root | Only root allowed | Only root login | No access to root shell | Only root allowed |

The malware used the **Telnet** port (23) to attempt to connect and gain root access by brute forcing default login credentials. Telnet is extremely insecure and transfers login credentials in plain text with its connections. It also offers the ability to perform important actions or modifications to the operating system, or access to root files that should be unmodified or tampered with. SSH is a more secure protocol to use, and although it is vulnerable in newer strands of the malware becoming available, a strong set of login credentials can make the brute forcing of an attacker extremely difficult and time depleting. Disabling Telnet, however, can be made extremely difficult as devices running the likes of BusyBox cannot support SSH without having to recompile the system. If Telnet, however, is essential, changing the port number it runs on to for example '2330' will be beneficial as the malware's loader will a large majority of the time scan port 23. Our suggested counter measure is that the Telnet protocol should be disabled and SSH used in preference if possible. Change Telnet port if the protocol is essential to the device.

Some of the devices were running the **HTTP protocol** on port **80**. HTTP creates what is deemed an insecure connection which is prone to interception and eavesdropping by attackers. HTTPS however, on port 443, offers a secure encrypted connection using SSL certificates which allows data sent over the web to be encrypted and secured from anybody unauthorised or not the recipient to view the data. Our suggested counter measure is to disable the HTTP on port 80 and use HTTPS on port 443. By reviewing the code we came to the conclusion that there is a rare occurrence when the mirai malware would connect the device to the IP address **5.206.225.96** on port **23** Telnet. This address hosts a game website, however blocking this IP is a good form of protection and the device should not be allowed to access it. Our suggested counter measure is to block access to the aforementioned IP address.

In the release of the Mirai's source code online by the claimed author Anna-senpai [28], a text file was in the post which clarified how some of the source codes processes function. It explained how the **port 48101** was used to prevent multiple instances of the bots running together and therefore was partially needed for the malware to operate correctly. Blocking this port can help secure the possibility of the malware infecting the device. If the IoT device runs **BusyBox**, the commands issued by the malware are executed in the BusyBox environment. A method of securing and preventing this is to secure the BusyBox execution to be run by a user that is specified by the device owner. Therefore, the malware will not have permission to execute its commands and continue its spreading of the infection.

IP cameras are beginning to integrate smart phone applications into their devices where the users can view the live camera feed directly from the app. Another feature of this app is to send email alerts via **SMTP** to the user's email address to alert them of something such as a trigger of the motion detectors. As counter measure we propose to disable the SMTP protocol which is normally run through port 25 but this might need to be checked as it can be run on other ports. **FTP** servers are normally enabled by default on IoT devices and can be used to transfer files to and from the device. This can result in dangerous files being moved onto the device and should be disabled to prevent this. **TFTP** can be used by the malware if the device does not support the *wget* tool and this should also be disabled.

To automate the process of hardening the devices, a script was developed to be run on the device which made the necessary essential changes that the devices allowed. Limitations were presented in the design of the script as devices running Linux with BusyBox had limited functionality and did not allow all the proposed defensive measures to be implemented. We also devleopped a piece of code written in the programming language C which is the same language the Mirai's malware uses. It is designed to locate the processes in the */proc* folder running on the device which are using the three folders */root/*, */tmp/* and */var/tmp/*. The scrpit will attempt to narrow and refine the search until only the processes the Mirai malware is using are left and it will use the kill function to end the process. This in turn will stop the infection and any communication with the CNC server. We also used another script developped by Frank et al. [29], which introduced another approach to prevent infection of the device via the malware. Its function was to create a temporary BusyBox wrapper, which when the known methods of the Mirai's malware were used, it would execute them onto the BusyBox wrapper and not directly onto the device.

Soon after we were done with the implementation of the proposed defensive measures, we attempted to infect the devices once more using our initial methodology. As expected, this time attempts did not have any success.

## V. CONCLUSIONS AND FUTURE WORK

As the complexity and functionality of IoT devices increase, it won't just be consumers that adopt them. Cities and companies will also proceed to adopt the technology to make everything smart and connected while saving money and time. This worldwide adoption, however, provides a fertile feeding ground for attackers to prey on and with more and more devices connected worldwide, it will only continue to facilitate their hunger to design attacks. A lot of malware targeting IoT devices have been built and among them, Mirai is very popular due to infecting millions of devices.

The scope of our work was to test how secure and especially how resistant everyday IoT devices can be against the Mirai malware. We built our testbed environment using three low cost IoT devices and one virtualised, simulated, device. Our experiments concluded that three out of the four devices were vulnerable to the Mirai malware and became infected. This proved that the factory security configurations are not sufficient and at an acceptable level for consumers and leaves their devices exposed by default.

After evaluating the success rate of our attacks, we came up with a range of defensive measures that would harden the IoT devices against the Mirai malware. Meeting our expectations, the security configurations proved to work, as none of the devices could be infected again using our initial attack methodology. We hope that the proposed solutions can be used by companies in the original design of devices and their firmware to ensure the security implemented is at an acceptable level for consumers.

If the Mirai botnet was able to bring down a number of popular websites, the potential for even more dangerous forms of attacks will become a serious security concern. As the size of botnets grows and the attacks become more sophisticated, the IoT devices could be weaponised and used to take down entire power grids. Acknowledging the fact that the line of defence needs to begin at the end point, i.e. the physical devices, we plan to test a wider range of devices as our future work. Additionally, our end goal is to create an automated software tool that will evaluate their vulnerabilities of the aforementioned devices against known attack vectors, thus supporting both the producers and the consumers.